\documentclass{article}
\usepackage[T1]{fontenc} 
\usepackage[utf8]{inputenc} 
\usepackage{amsmath,cite,url}
\usepackage{graphicx}
\graphicspath{{fig/}}
\usepackage[lbd]{ismir}

\usepackage{color}
\usepackage{xcolor}
\definecolor{earthyellow}{rgb}{0.82,0.70,0.45}   
\newcommand{\cw}[1]{#1}
\newcommand{\amy}[1]{#1}
\newcommand{\icw}[1]{#1}
\usepackage{wrapfig}

\usepackage[firstpage]{draftwatermark}
\definecolor{lightgray}{rgb}{0.9,0.9,0.9}
\definecolor{darkgray}{rgb}{0.4,0.4,0.4}
\SetWatermarkFontSize{12pt}
\SetWatermarkScale{1.1}
\SetWatermarkAngle{90}
\SetWatermarkHorCenter{202mm}
\SetWatermarkVerCenter{170mm}
\SetWatermarkColor{darkgray}
\SetWatermarkText{Late-Breaking / Demo Session Extended Abstract, ISMIR 2025 Conference}

\title{Timed text extraction from Taiwanese Kua-á-hì TV series}

\multauthor
{Tzu-Hung Huang$^{1,4}$ \hspace{1cm} Yun-En Tsai$^1$ \hspace{1cm} Yun-Ning Hung$^3$} { \bfseries{Chih-Wei Wu$^2$ \hspace{1cm} I-Chieh Wei$^5$ \hspace{1cm} Li Su$^1$}\\
$^1$ Academia Sinica, Taiwan \hspace{1cm} 
$^2$ Independent Researcher \hspace{1cm} 
$^3$  Music AI, USA \hspace{1cm} \\
$^4$ National Taiwan University, Taiwan \hspace{1cm} $^5$ University of Aukland, New Zealand\\
{\tt\small b11902023@ntu.edu.tw, lisu@iis.sinica.edu.tw}
}

\def\authorname{T.-H. Huang, Y.-E. Tsai, Y.-N. Hung, C.-W. Wu, I-C. Wei, and L. Su}

\begin{document}

\maketitle
\begin{abstract}

\amy{Taiwanese opera (Kua-á-hì), a major form of local theatrical tradition, underwent extensive television adaptation notably by pioneers like Iûnn Lē-hua.}
\cw{These videos, while potentially valuable for in-depth studies of Taiwanese opera, often \amy{have low quality and} require substantial manual effort during data preparation.} 
\amy{To streamline this process, we developed an interactive system for real-time OCR correction and a two-step approach integrating OCR-driven segmentation with Speech and Music Activity Detection (SMAD) to efficiently identify vocal segments from archival episodes with high precision.}
\cw{The resulting dataset, consisting of vocal segments and corresponding lyrics, can potentially supports various MIR tasks such as lyrics identification and tune retrieval.} 
Code is available at \url{https://github.com/z-huang/ocr-subtitle-editor}.

\end{abstract}
\section{Introduction}\label{sec:introduction}
Kua-á-hì \cite{hsu2010living} emerged in the early 20th century as a form of indigenous Taiwanese opera performed in the Taiwanese language (Tâi-gí). 
\cw{This theatrical form combining acting and singing in a cohesive dramatic narration, was}
traditionally performed during religious celebrations and temple festivals to honor deities. In the 1980s, with the widespread adoption of television broadcasting, Kua-á-hì began to be extensively adapted into television drama series. Among these adaptations, the master Iûnn Lē-hua, who joined Taiwan Television Enterprise for more than a decade, 
produced a considerable body of work and is widely regarded as the pioneer of televised Kua-á-hì. These productions are now available online, offering valuable resources for understanding and exploring Kua-á-hì through Music Information Retrieval (MIR) approaches.

\cw{These videos preserve all the essential components for studying Taiwanese opera, including audio, video, and dialogue/lyrics via burned-in subtitles.} 
However, due to the age and the poor quality of videos, conventional OCR technologies cannot achieve satisfactory accuracy, necessitating time-consuming manual corrections.
\cw{To efficiently prepare the data for various MIR tasks and explorations, we developed an interactive OCR tool for subtitle extraction and correction. Additionally, we proposed a two-step workflow that integrates OCR-driven segmentation with a SMAD algorithm to prepare a Taiwanese opera dataset with vocal segments and their corresponding lyrics. The detailed process will be elaborated in the following sections.}

\begin{figure}[t]
 \centerline{
 \includegraphics[width=\columnwidth, trim={1cm 2cm 1cm 0cm},clip]{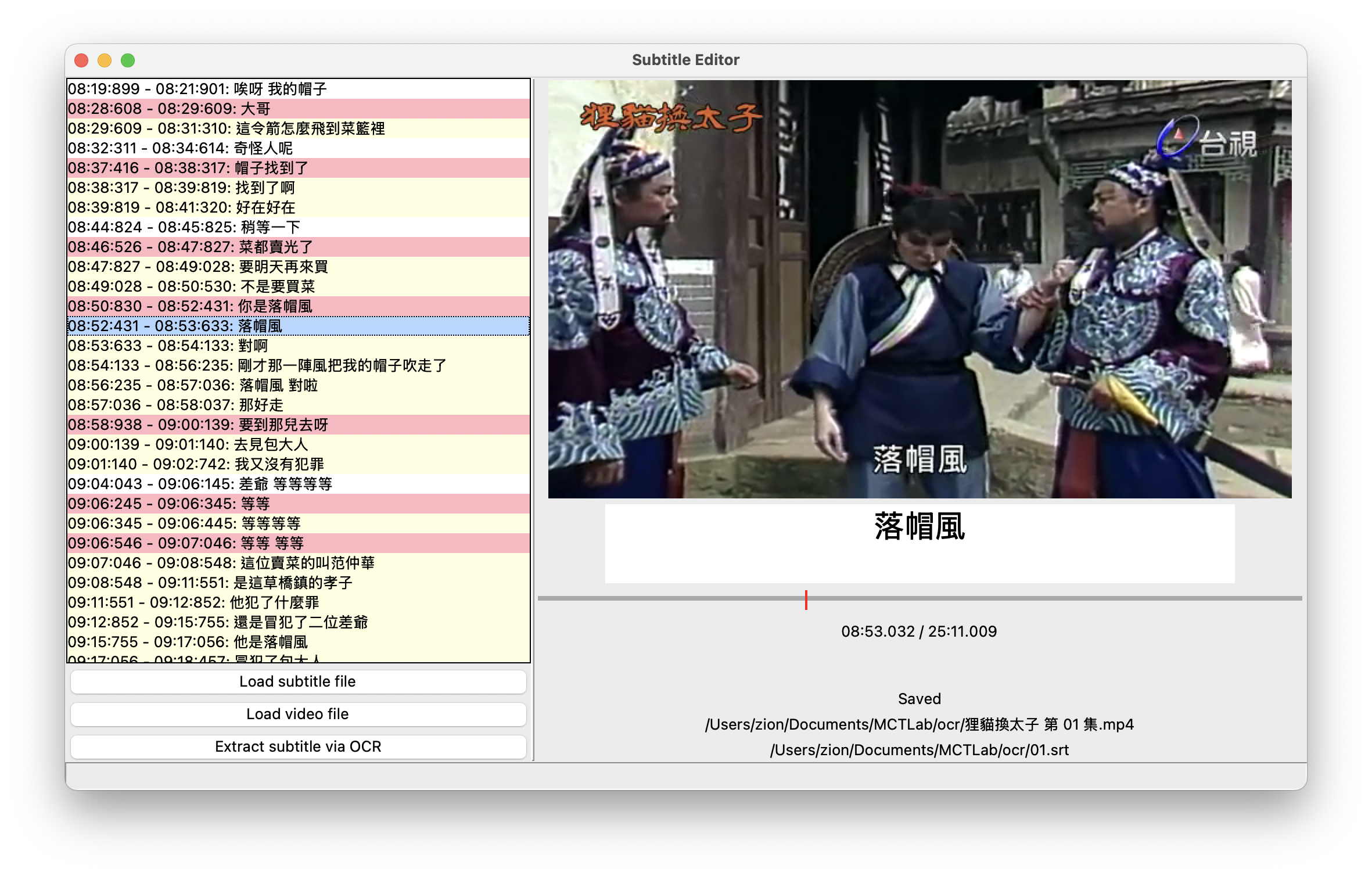}}
 \caption{Graphical user interface of the subtitle editor.}
 \label{fig:editor_gui}
\end{figure}

\begin{figure*}[t]
 \centering
 \includegraphics[width=\textwidth, trim={0cm 0.3cm 0cm 0.1cm},clip]{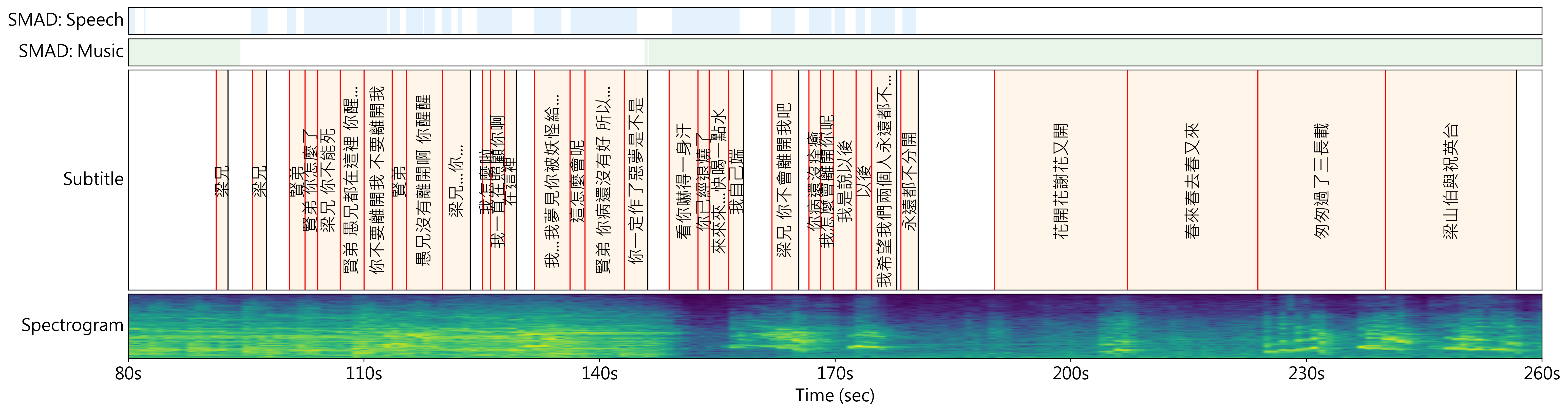}
 \caption{\amy{Spectrogram and estimated subtitle, music and speech of one episode in the dataset.}}
 \label{fig:enter-label}
\end{figure*}

\section{\amy{Subtitle Extraction}}\label{sec:method}

\subsection{Text Extraction via OCR}

Our subtitle extraction pipeline consists of three stages: text extraction via OCR, temporal merging, and manual correction through a Graphical User Interface (GUI). 
To support traditional Chinese recognition with high accuracy and cost efficiency, we selected EasyOCR~\cite{easyocr} as the core engine of our system. EasyOCR provides both bounding box coordinates of detected text and confidence scores, which are essential for downstream processing.
\cw{To balance computational efficiency with temporal accuracy, we perform OCR once every 0.1 second. Additionally, OCR processing is restricted to the region where subtitles typically appear, reducing interference from other on-screen elements (e.g., TV station's logo).}
We retain 
those subtitle strings detected by EasyOCR whose confidence score exceeds 0.01, then sort them from left to right and concatenate them with spaces.


The output from the \icw{OCR} 
stage contained substantial noise, including typographical errors, false positives, and intermittent character recognition across frames. 
\cw{To clean and consolidate the text sequences, we implemented the following temporal merging mechanism.}
First, only text sequences with durations longer than a predefined threshold $t$ are retained \cw{to reduce false positives.}
\cw{Additionally, we maintain a list of unlikely subtitle characters and remove them from the text.}
\cw{To deduplicate two adjacent text sequences, a similarity check based on edit distance~\cite{levenshtein1966, wagner1974} and a threshold was implemented.}
Specifically, for short strings ($\le$ 2 characters), we require exact matches; for medium-length strings ($\le$ 6 characters), we accept an edit distance of 1; for longer strings ($>$ 6 characters), a maximum edit distance of 2 is allowed. \cw{If two similar text sequences occur in succession, we keep the one with a higher confidence score.}
Furthermore, when the base text (excluding minor differences such as ellipses) is the same, we update the confidence score to the maximum and retain the longer version of the text. 
If a new, dissimilar text appears after the previous text has lasted beyond $t$, the prior subtitle segment is finalized and stored. 
\cw{This frame-wise process produces} a temporally stable sequence of subtitles with 
improved consistency.
\cw{The current set of parameters is selected to prioritize the retention of true positive segments, even at the expense of some false positives, which are filtered out in the subsequent stage.}

\subsection{Manual Correction through GUI Interface}

\cw{To support efficient manual correction in the last stage of data refinement, we developed a custom subtitle editing interface using Python tkinter~\cite{tkinter}.}
\cw{As shown in Figure~\ref{fig:editor_gui}, the interface, featuring keyboard shortcuts, is lightweight yet optimized for rapid editing.}
The interface presents a dual-pane layout: the left pane lists subtitle segments, while the right pane shows the corresponding video frame, a text editor, and a timeline control. 
\cw{By pressing the \texttt{e} key and \texttt{Esc} key, the user can quickly enter and exit edit mode while viewing the content.}
To further streamline the merging process,
temporally adjacent subtitles are marked with a yellow background, and the beginning of each new sequence is indicated with a red background. These visual cues help the user to quickly identify and merge subtitle segments.
\cw{Note that all the functionalities of the previous two stages can be invoked using the "Extract Subtitle via OCR" button, allowing a seamless transition across all three stages within the same tool.}

\section{Dataset Preparation}\label{sec:datasetpreparation}

The vocal content of Kua-á-hì contains both speech and singing components, with our analysis focusing specifically on singing. To isolate the singing segments, we employed a two-stage approach that combines OCR technology and a SMAD algorithm to \cw{speed up the process.}
\amy{As a preprocessing method to isolate the core element of Taiwanese opera performance, a music source separation algorithm \cite{moises} 
was utilized to extract the vocal stem from the original mixtures. Since our preliminary analysis reveals that singing segments exhibit longer subtitle display durations (0.9 seconds per character) compared to speech segments (0.4 seconds per character) due to melodic elongation in vocal delivery \cite{doi:10.1126/sciadv.adm9797_fix}, we used OCR-generated subtitle files to extract potential music segments, where a minimum of four consecutive characters with a display duration exceeding 0.4 seconds were classified as valid singing components.}
Subsequently, the SMAD results were integrated to calculate temporal overlaps between OCR-derived segments and regions classified as music content, minimizing false positives such as slow speech patterns and instrumental sections. Figure \ref{fig:enter-label} illustrates the estimated SMAD and subtitle information.

\icw{The combined OCR + SMAD pipeline greatly streamlines segment identification. Starting from 220 episodes (94.28 h), OCR-based segmentation produced 1,798 candidate segments, cutting the material that needed human inspection to 45.73 h—a 51.5\% reduction. SMAD filtering narrowed this set to 1,400 segments(42.67 h). A final auditory check confirmed 841 of them(35.68 h) as genuine singing passages lasting 22.47 – 1,352.10 s. Annotators now review only 6.4 segments per episode on average, while the overall pipeline attains a precision of 60.1\%.}

\section{Conclusion}\label{sec:conclusion}
\cw{This study presents an interactive system for extracting subtitles from low-quality videos and its direct application for preparing a dataset of Kua-á-hì. With the proposed approach, we build a dataset from archival television materials, allowing future explorations of MIR tasks like lyrics identification and tune retrieval.}
Beyond Kua-á-hì, this framework shows potential for processing medium- to low-quality historical television content, 
facilitating broader applications in digital humanities and cultural preservation studies.



\bibliography{ISMIRtemplate}

\end{document}